\documentclass[final]{aipproc}

\layoutstyle{6x9}

\makeatletter
\newcommand\erfc{\mathop{\operator@font erfc}\nolimits}
\def\slashchar#1{\setbox0=\hbox{$#1$}
   \dimen0=\wd0 \setbox1=\hbox{/} \dimen1=\wd1
   \ifdim\dimen0>\dimen1 \rlap{\hbox to \dimen0{\hfil/\hfil}} #1
   \else  \rlap{\hbox to \dimen1{\hfil$#1$\hfil}} / \fi}

\makeatother

\begin{document}

\title{Generalized parton distributions of the pion\footnote{Talk presented by WB at
{\em SCADRON 70 Workshop on ``Scalar Mesons and Related Topics''}, Lisbon, 11-16 February 2008}}



\author{Wojciech Broniowski}{
  address={The H. Niewodnicza\'nski Institute of Nuclear Physics PAN, PL-31342 Krak\'ow, Poland and 
           Institute of Physics, Jan Kochanowski University, PL-25406~Kielce, Poland}
}

\author{Enrique Ruiz Arriola}{
  address={Departamento de F\'{\i}sica
At\'omica, Molecular y Nuclear, Universidad de Granada, E-18071 Granada, Spain}
}

\author{Krzysztof~Golec-Biernat}
 {address={The
H. Niewodnicza\'nski Institute of Nuclear Physics PAN, PL-31342 Krak\'ow,
Poland, and Institute of Physics, Rzesz\`ow
University, PL-35959 Rzesz\'ow, Poland}
}

\begin{abstract}
Generalized Parton Distributions of the pion are evaluated in chiral
quark models with the help of double distributions. As a result the
polynomiality conditions are automatically satisfied. In addition, 
positivity constraints,
proper normalization and support, sum rules, and soft pion
theorems are fulfilled. We obtain explicit expressions holding at the
low-energy quark-model scale, which exhibit no factorization in the
$t$-dependence. The crucial QCD evolution of the quark-model
distributions is carried out up to experimental or lattice scales. The
obtained results for the Parton Distribution Function and the Parton
Distribution Amplitude describe the available experimental
and lattice data, confirming that the quark-model scale is low, around
320~MeV.
\end{abstract}

\maketitle

Generalized Parton Distributions (GPD's) carry ``tomographic''
information on the partonic structure of hadrons (for reviews see
e.g.~\cite{Ji:1998pc,Radyushkin:2000uy,Goeke:2001tz,%
Diehl:2003ny,Ji:2004gf,Belitsky:2005qn,Feldmann:2007zz,Boffi:2007yc}).
In this talk we present our recent calculation of the GPD's of the pion
in the framework of chiral quark models \cite{Broniowski:2007si},
which extends the previous calculations of PDF's
~\cite{Davidson:1994uv,RuizArriola:2001rr,Davidson:2001cc}, PDA's
\cite{RuizArriola:2002bp, RuizArriola:2002wr}, and GPD in the impact
parameter space \cite{Broniowski:2003rp}. Recently, the Transition
Distribution Amplitudes (TDA) \cite{Pire:2004ie,Pire:2005ax} have also
been evaluated in the same framework~\cite{Broniowski:2007fs}.  Other
quark-model calculations of GPD's and related quantities have been
reported in
Refs.~\cite{Polyakov:1998td,Polyakov:1999gs,Anikin:2000th,Anikin:2000sb,Tiburzi:2002kr,Tiburzi:2002tq,%
Theussl:2002xp,Tiburzi:2002kr,Praszalowicz:2003pr,Bzdak:2003qe,Noguera:2005cc,Courtoy:2007vy,%
Kotko:2008gy}.

Chiral quark models yield parton distributions at a {\it given} low
energy scale $Q_0$.  The result for a quantity $F$ is matched to QCD
order by order in the twist expansion, $n$, hence $F_n (x) |_{\rm
Model} = F_n ( x, \alpha (Q_0^2 ) )|_{\rm QCD}$. Then the functions
$F_n$ are evolved to higher scales $Q$.  It turns out that in order to
describe the available pion phenomenology the initial scale $Q_0$ in
the considered quark models must be very
low~\cite{Davidson:1994uv,Davidson:2001cc,RuizArriola:2002bp,RuizArriola:2002wr};
matching the momentum fraction carried by the valence quark at $Q^2 =
4 {\rm GeV}^2 $ to $47 \%$~\cite{Sutton:1991ay,Capitani:2005jp} yields
\begin{eqnarray}
Q_0 = 313_{-10}^{+20} {\rm MeV}, \label{Q0}
\end{eqnarray}
with $\Lambda_{\rm QCD}=226$~MeV and three flavors.
At such a low scale ${\alpha(Q^2_0)}/({2\pi})=0.34$, which
makes the evolution very fast for the scales close to the initial value. 

\begin{figure}[tb]
\includegraphics[width=0.4\textwidth]{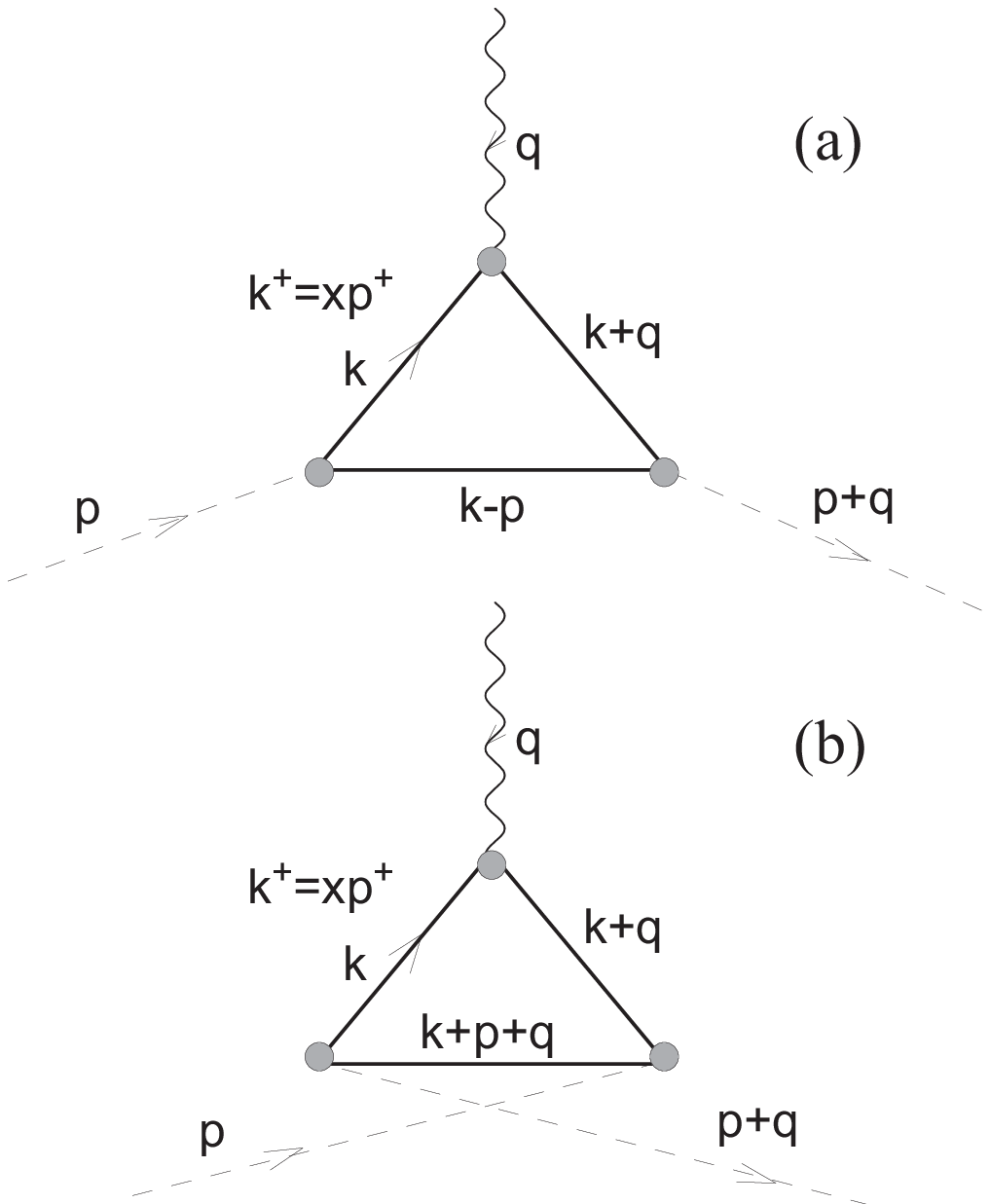} 
\includegraphics[width=0.27\textwidth]{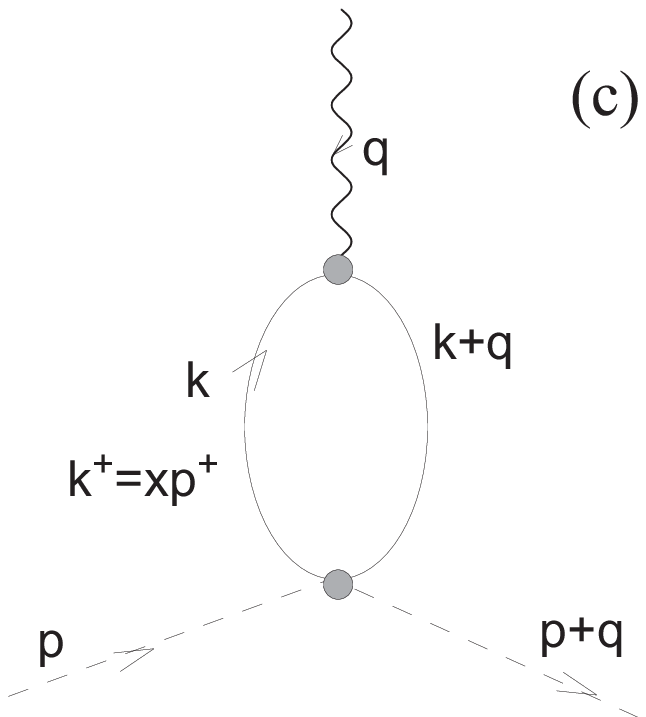} 
\caption{The direct (a), crossed (b), and contact (c) Feynman diagrams for the
quark-model evaluation of the GPD's of the pion.
\label{fig:diag}}
\end{figure}

The kinematics of the process and the assignment of momenta (in the
asymmetric notation) is displayed in Fig.~\ref{fig:diag}, representing the 
large-$N_c$ quark-model evaluation of GPD's. We adopt
the standard notation $p^2=m_\pi^2$,  $q^2=-2p\cdot q=t$, $q \cdot n=-\zeta$.
The leading-twist GPD of the pion is defined as
\begin{eqnarray}
 {\cal H}^{ab}(x,\zeta,t) = \int \frac{dz^-}{4\pi} e^{i x p^+ z^-}
\label{defGPD}  \left . \langle \pi^b (p+q) | \bar
\psi (0) \gamma \cdot n \,T\, \psi (z) | \pi^a (p) \rangle \right
|_{z^+=0,z^\perp=0},
\end{eqnarray} 
where $a$ and $b$ are isospin indices for the pion,
$T$ is the isospin matrix equal $1$ for the isoscalar and $\tau_3$ for
the isovector case, $n$ is the null vector, and $z$ is the light-cone
coordinate.  In the symmetric notation one introduces
$\xi= \frac{\zeta}{2 - \zeta}$ and  $X = \frac{x - \zeta/2}{1 -
\zeta/2}$. The following sum rules hold on general grounds:
\begin{eqnarray}
\int_{-1}^1 \!\!\!\!\! dX\, {\cal H}^{I=1}(X,\xi,t) = 2 F_V(t), \label{norm} \;\;\;
\int_{-1}^1 \!\!\!\!\! dX\,X \, {\cal H}^{I=0}(X,\xi,t) = \theta_2(t)-\xi^2 \theta_1(t), 
\end{eqnarray}
where $F_V(t)$ is the electromagnetic form factor, while $\theta_1(t)$
and $\theta_2(t)$ are the gravitational form factors of the pion. 
Finally, for $X \ge 0$ the equality  ${\cal H}^{I=0,1}(X,0,0) = q(X)$
relates the GPD's to the the pion's parton
distribution function (PDF).
The {\em polynomiality} conditions~\cite{Ji:1998pc,Radyushkin:2000uy}
and the positivity bound~\cite{Pobylitsa:2001nt} are satisfied in our approach.

\begin{figure}[tb]
\includegraphics[width=1.07\textwidth]{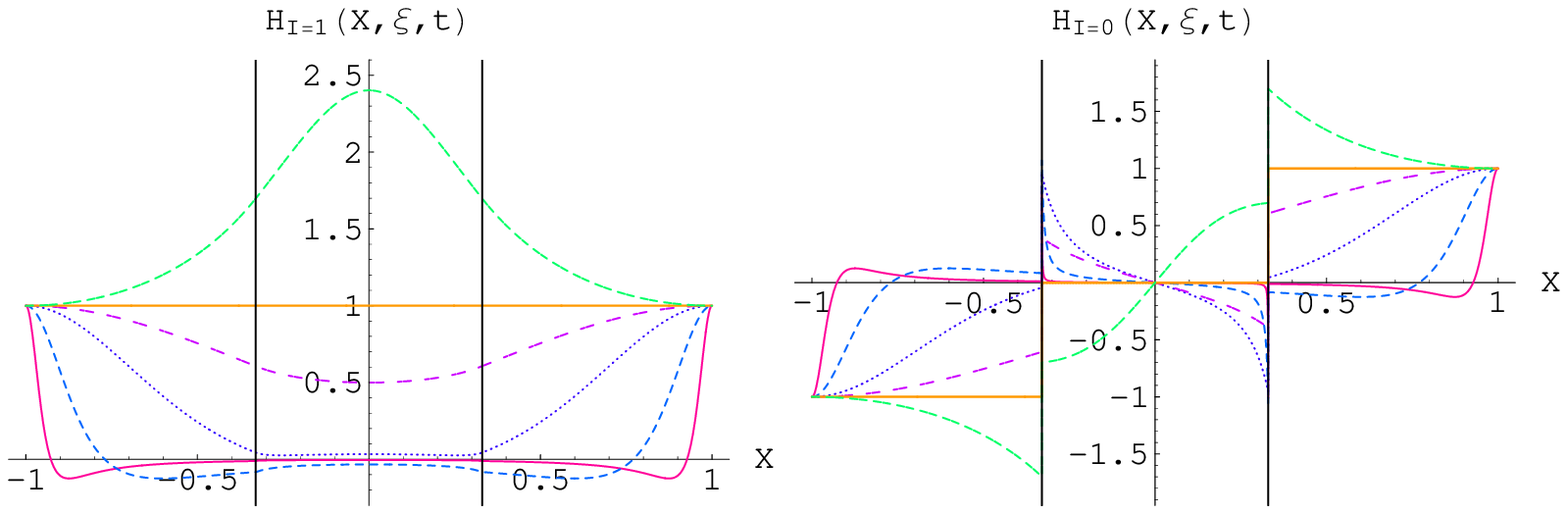}
\caption{The GPD's of the pion in SQM, $\xi=1/3$, \mbox{$t=0.2,0,-0.2,-1,-10,-100~{\rm GeV}^2$} from top
to bottom (at $x=0.9$). \label{fig:H01}}
\end{figure}

We work for simplicity in the chiral limit, $m_\pi=0$.
Two quark models are considered: the Spectral Quark Model (SQM)~\cite{RuizArriola:2003bs}
and the NJL model. SQM implements the vector-meson dominance, predicting the form factors
\begin{eqnarray}
F_V^{\rm SQM}(t)=\frac{M_V^2}{M_V^2-t}, \;\;\;
\theta_1^{\rm SQM}(t)=\theta^{\rm SQM}_2(t)=\frac{M_V^2}{t} \log \left
( \frac{M_V^2}{M_V^2-t} \right ).  \label{ffSQMg}
\end{eqnarray}
The explicit results for the full GPD's 
have been provided in Ref.~\cite{Broniowski:2007si}. 
Importantly, their form does not exhibit a factorized $t$-dependence.
A sample result for $\xi=1/3$ and several values of $t$
is shown in Fig.~\ref{fig:H01}. For the NJL model the 
results are qualitatively the same.
For the case of $t=0$ the GPD's simplify to the
well-know \cite{Polyakov:1999gs,Theussl:2002xp} step-function results
\begin{eqnarray}
{\cal H}_{I=0}(x,\zeta,0) &=& \theta[(1 - x)(x - \zeta )] - 
\theta[-x(x + 1 - \zeta )], \nonumber \\ {\cal H}_{I=1}(x,\zeta,0)&=&\theta[(1-x)(x+1-\zeta)].  \label{zerot}
\end{eqnarray}
Another simple case is in SQM for $\zeta=0$ and any value of $t$ \cite{Broniowski:2003rp}
\begin{eqnarray}
{\cal H}_{q}(x,0,t) =\frac{M_V^2 \left(M_V^2+t
(x-1)^2\right)}{\left(M_V^2-t (x-1)^2\right)^2}.
\end{eqnarray}

\begin{figure}[tb]
\includegraphics[width=0.5\textwidth]{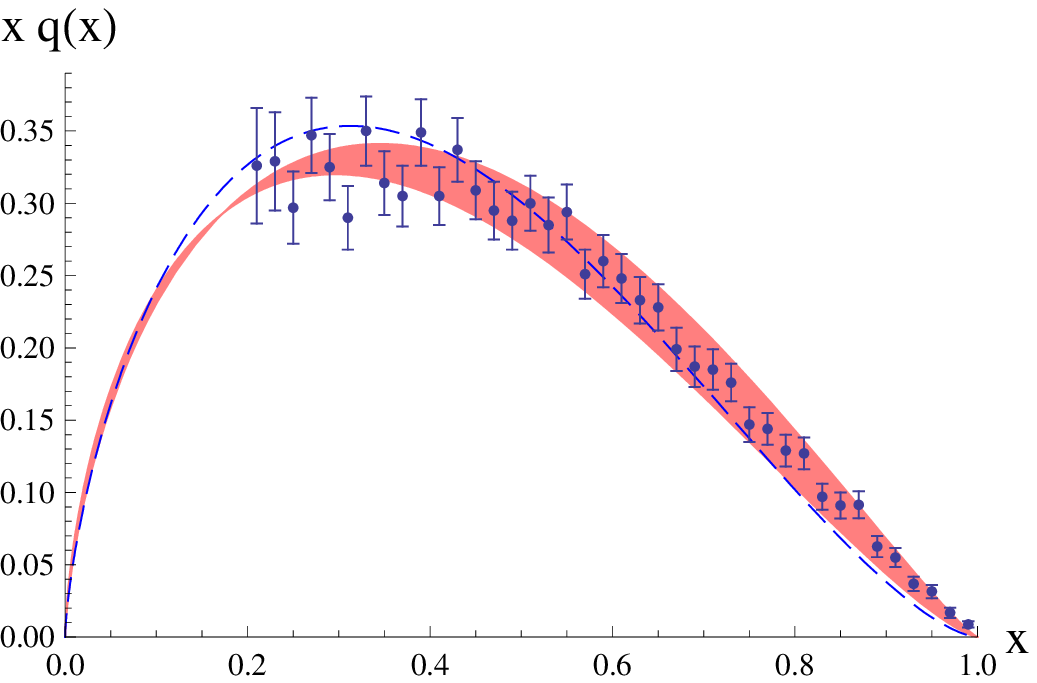}
\includegraphics[angle=0,width=0.45\textwidth]{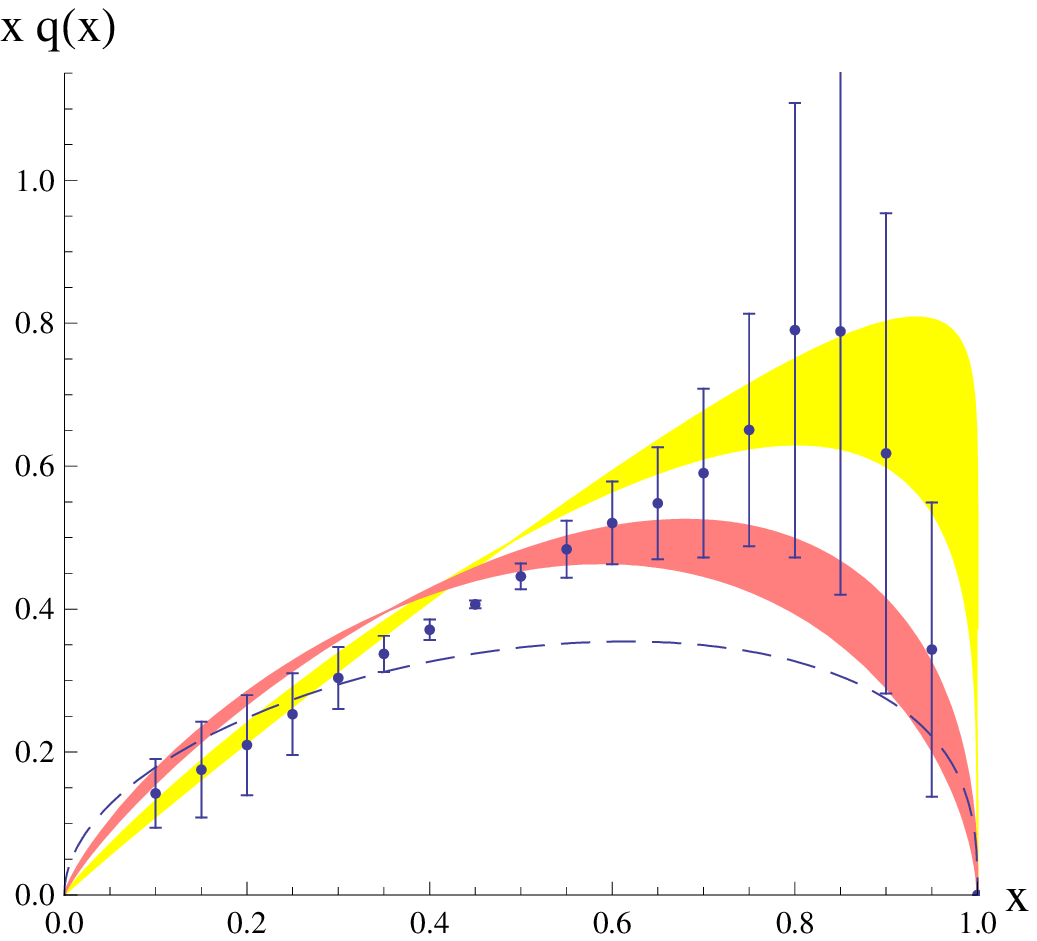}
\caption{Left: the quark model valence
parton distribution (PDF) of the pion for a single quark (either $u$
or $\bar d$ for $\pi^+$) evolved to the scale of $Q=4$~GeV (band). The
width of the band indicates the uncertainty in the initial scale
$Q_0$. The data points come from the analysis of the E615 experiment \cite{Conway:1989fs}. The
dashed line shows the reanalysis of the original data from
Ref.~\cite{Wijesooriya:2005ir}. Right: the quark-model prediction for PDF 
evolved to the scale $Q=0.5~{\rm
GeV}$ (darker band) and $Q=0.35~{\rm GeV}$ (lighter band). The transverse-lattice 
data come from Ref.~\cite{Dalley:2002nj} and correspond to the scale $\sim
0.5$~GeV. The line shows the GRS parameterization
at $Q=0.5~{\rm GeV}$. \label{e615}}
\end{figure}

For the QCD evolution we use the leading-order
ERBL-DGLAP equations with three flavors.  
In the left panel of Fig.~\ref{e615} we confront
the result for $x q(x,Q)$ at the scale $Q=2$~GeV with the data at this 
scale from the E615 Drell-Yan experiment
\cite{Conway:1989fs}. We note agreement between the model and the data.
In the right panel of Fig~\ref{e615} we compare 
our results to the data from lattices \cite{Dalley:2002nj}. We take the liberty of moving the scale, as its
determination on the lattice is not very precise.  As we see, the
agreement is qualitatively good if one considers the uncertainties of
the data, especially when the lower scale is used. 

\begin{figure}[tb]
%%\vspace{-13mm}
%%\begin{center}
\includegraphics[angle=0,width=0.47\textwidth]{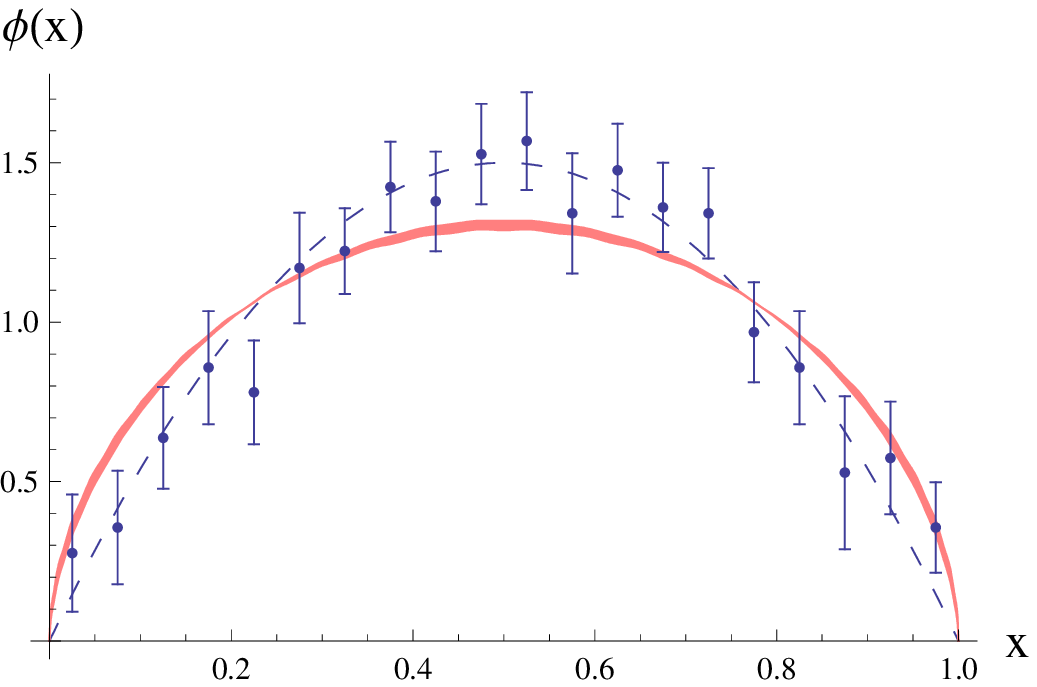}
\includegraphics[angle=0,width=0.47\textwidth]{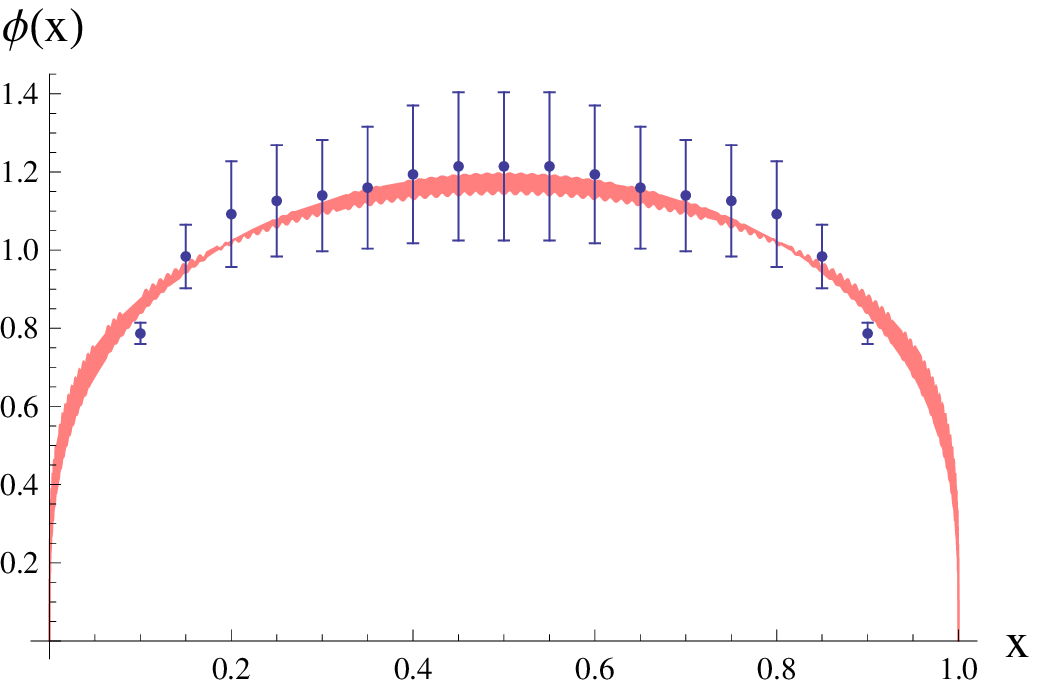}
%%\end{center}
\vspace{-1mm}
\caption{Left: the quark-model prediction for the pion distribution amplitude (PDA)
evolved to the scale $Q=2~{\rm GeV}$ (band) and compared to the E791 di-jet
measurement~\cite{Aitala:2000hb} after proper normalization of the data. The width of the band indicates the 
uncertainty in $Q_0$. We
also show the the asymptotic PDA, $\phi(x,\infty)=6x(1-x)$ (dashed line). 
Right: the same compared to the transverse lattice data~\cite{Dalley:2002nj}, corresponding to the scale 
$\sim 0.5$~GeV. 
\label{fig:pda-evol}}
\end{figure}

PDA's have been intensely studied in the past in several contexts (see
Ref.~\cite{Bakulev:2007jv} for a brief review). At the quark model scale $Q_0$ the
PDA of the pion \cite{RuizArriola:2002bp}, which can be related to the
isovector GPD through the soft pion theorem \cite{Polyakov:1998ze} is
$\phi(x;Q_0) = 1$ \cite{RuizArriola:2002bp}.  The evolved PDA is shown in
Fig.~\ref{fig:pda-evol}, where it is compared to the E791 di-jet
measurement~\cite{Aitala:2000hb} and to lattice calculations \cite{Dalley:2002nj}. 
Again, good agreement is observed.

For the case of general kinematics, 
the explicit form of the LO QCD evolution equations for the GPD's can
be found in
\cite{Mueller:1998fv,Ji:1996nm,Radyushkin:1997ki,Blumlein:1997pi,GolecBiernat:1998ja,Kivel:1999wa,Kivel:1999sk}.
In this paper we solve them with the numerical method developed in
\cite{GolecBiernat:1998ja}, based on the Chebyshev polynomial
expansion. 
\begin{figure}[tb]
\includegraphics[width=0.69\textwidth]{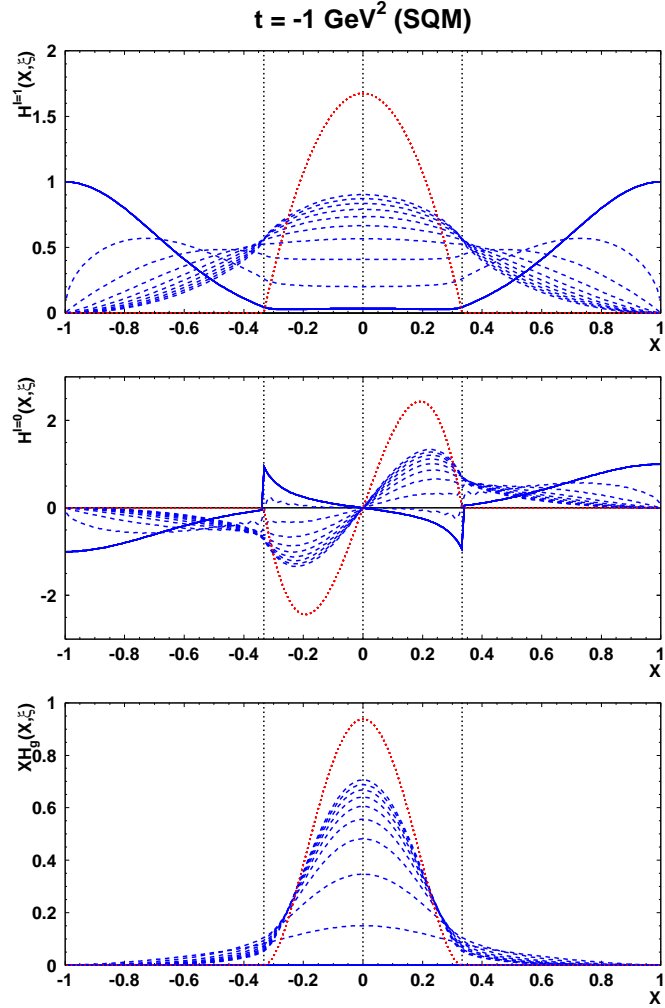}
\caption{Non-singlet (top), singlet (middle), and gluon (bottom) GPD's of the pion for $\xi=1/3$, evolved 
from the SQM condition up to $Q^2=0.1,1,10,10^2,\dots,10^8$~GeV${}^2$. Higher $Q^2$ 
gives higher magnitude of the curves in the ERBL region. \label{resQ}}
\end{figure}
The results of the LO evolution from the SQM initial condition for
$\xi=1/3$ at the scale $Q_0$ to subsequent values of $Q$ are shown in
Figs.~\ref{resQ}. The evolution is fastest at low values of $Q$, where
the coupling constant is large, and it immediately pulls down the
end-point values to zero. Then, the strength gradually drifts from the
DGLAP regions to the ERBL region.  The approach towards the asymptotic
form is very slow, with the tails in the DGLAP region present. The
highest $Q^2$ displayed in the figure is $10^8$~GeV${}^2$ and the
asymptotic form is reached at ``cosmologically'' large values of $Q$,
which are never achieved experimentally.  The results for the NJL
model are very similar to the case of SQM.

In conclusion, we remark that our calculation provides a link between
the non-perturbative soft-energy physics in terms of matrix element of
operators and the high-energy processes as deduced from perturbative
QCD evolution.  The overall agreement with the pionic data from
experiments and lattices, available for the PDF and PDA, is very
reasonable, supporting the presented methodology.

\bigskip
Supported by Polish Ministry of Science and Higher
Education grant N202~034~32/0918, Spanish DGI and FEDER
funds with grant FIS2005-00810, Junta de Andaluc{\'\i}a grant
FQM225-05, and EU Integrated Infrastructure Initiative Hadron Physics
Project contract RII3-CT-2004-506078.

%\bibliographystyle{aipproc}   % if natbib is available
%\bibliography{gpd}

\end{document}